\documentclass[14pt,a4paper]{extarticle}

\usepackage{graphicx}
\usepackage{epsfig}
\usepackage{bm}

\usepackage{amsmath,amssymb,amsfonts}
\usepackage{comment}

\title{Relativistic runaway electron breakdown and Terrestrial Gamma ray Flashes: GEANT4 simulation}

\author{Yu. A. Tsalkou\thanks{skytec@mail.ru}, V. V. Tikhomirov \thanks{vvtikh@mail.ru},
H. P. Marozava\thanks{ganna.marozava@gmail.com}
\\
\small
\textit{
Research Institute for Nuclear Problems, Belarusian State University} \\
\small
\textit{ 11 Bobruiskaya str., Minsk, 220050, Belarus}
}

\date{}

\begin{document}

\maketitle

\vspace{-1.0cm}
\begin{abstract}

Numerical simulation of a relativistic runaway electron breakdown in the upper atmosphere is performed using GEANT4 simulation toolkit.
General features of a relativistic runaway electron avalanche are reconstructed and properties of radiations accompanying breakdown are obtained.
It is demonstrated dependance of the high energy branch of photon spectra with respect to an altitude and shown what at the reasonable parameters hard photons have spectral index close to the observed value.

\end{abstract}

\section{Introduction}

Last decades were rich enough for numbers of the remarkable discoveries of different transient electromagnetic phenomena in the upper atmosphere. They covered wide region of the spectra from optical and even down to Extremely Low Frequencies -- so cold "Transient Luminous Events" (TLE) like "sprites",  "elves" and "jets" \cite{Sprites, ELF, TLE}, up to X and gamma rays -- so cold "Terrestrial Gamma Flashes" (TGF) \cite{1994Sci...264.1313F, 2005Sci...307.1085S}.
Additionally to the electromagnetic radiation the electron-positron beams were observed which are ejected from the Earth to the outer space \cite{el-pos}.
All these phenomena are related in some way to a thunderstorm's complex, they generated in a strong electric field in it and at lest some of them connected to the lightning flashes \cite{inan2006terrestrial}.
The energy level of some events is so huge that a question about a neutrons photoproduction is appeared \cite{babich2006_n-generation, TGF_neutron}.

Its widely known that an electric field in a thunderstorm's system never attains the value of usual air breakdown $E_{th}\simeq 2,16\, kV\,cm^{-1}$, thus lightning initiates mostly by streamers and leaders \cite{uman2011lightning}. Usual fields in thunderstorm clouds are  order of magnitude lower, but as was first mentioned by Wilson it is possible for cosmic ray electrons to be accelerated and even to runaway due to a dependence between energy $\varepsilon$ of particle and slowing-down force $F$ of the media \cite{Wilson}. Namely, in the nonrelativistic limit $\varepsilon \ll m c^2$ ionization loses decreases with energy as $F \sim \varepsilon^{-1}\ln\varepsilon $. In the opposite case $\varepsilon \gg m c^2$ it increases slowly as $F\sim \ln \gamma$, where $\gamma$ is Lorentz gamma-factor. It is clear that $F$ has a minimum $F_{min}$ and if an electron with energy $\varepsilon>\varepsilon_c$ where $\varepsilon_c\approx m c^2E_c/2E$ gets in the electric field $E>E_c = F_{min}/e$ then it become continuously accelerated,  i.e. runaway.
This process can be accompanied by an appearance of additional $\delta$-electrons and avalanche growth of runaway electrons leading to large populations of energetic electrons is possible what results in the relativistic runaway electron breakdown \cite{gurevich-94}.

Because of relatively low electric field responsible for producing   such phenomena, the relativistic runaway electron avalanche (RREA) is  considered as the basic effect for generating such transient process like TLE and TGF. Moreover, it seems to play crucial role in usual thunderstorm activity due to sufficient increasing of conductivity in the cloud \cite{gurevich-ufn}.

Despite the progress in the theoretical description of energetic phenomena in the upper atmosphere several observed features remain  unsolved. It is not clear how all optical and high energetical phenomena related to each other \cite{Dwyer_HEP_Atm12}. The subject of discussion is a source mechanism and location of TGF \cite{dwyer2008source}.

Recently one more problem appeared from sensitive measurements of
AGILE Team \cite{PRL_AGILE_11} which confidently detect an excess in the hard part of the TGF spectra for $\varepsilon_\gamma > 10 MeV$ unexplained  by usual models.

Thus a detailed picture of the formation and development of RREA is needed to clarify the reasons of such peculiarities.
The question about an influence of TGF radiation on artificial satellites and aircrafts rises up additionally.

In order to trace generation of RREA and accompanied radiation we perform modeling of such process using GEANT4 simulation toolkit pay attention to  basically initial stages of generation.

In the next section we will describe the general set up of the modeling, then we address to features of obtained spectra in the section \ref{hard-sp-toc} and finally in conclusion some results will be summarized.

\section{GEANT4 simulation of runaway breakdown}

We start from preliminary set up and consider as a modeling volume ("world volume" in GEANT4 terminology \cite{GEANT4})
an initially uniform cylindrical layer of atmosphere with height $\Delta h=2000$ m, radius $R=1000$ m,  located on $h=5000$ m height and with uniform electric field $E=2 E_c$ oriented down to the Earth what correspond to the usual properties in the thundercloud during positive cloud to ground (+CG) discharge which can produce quasistatic electric field \cite{gurevich-ufn}. Density, pressure and temperature in the modeling volume correspond to their values on respective height $h$.
The "Detector" will be a thin $\Delta h = 10$ cm slice of an air as wide as the world volume there particles characteristics will be collected and its position can be chosen anywhere inside the cylinder. Now it will be close to the top of the modeling cylinder thereby the formation and further development of a RREA at the scale of 2 km are possible to observe in all details.

\begin{figure}[h!]
\begin{center}
\includegraphics[width=4 in]{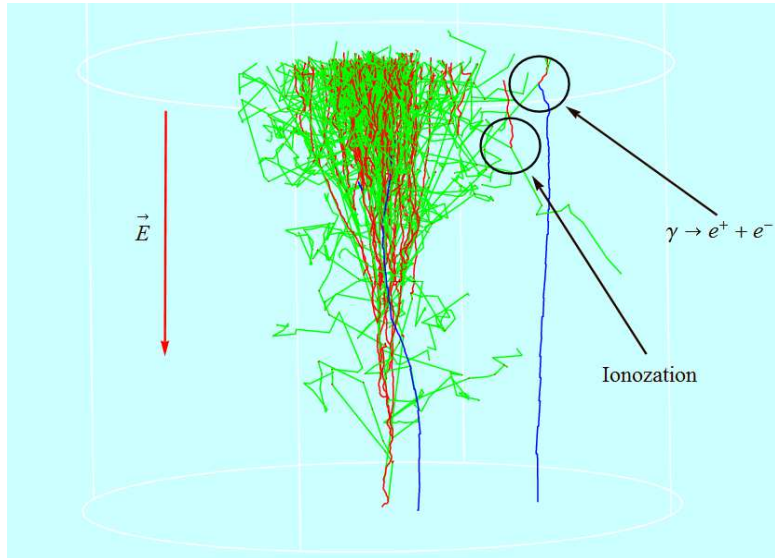}
\caption{Relativistic runaway electron avalanche. Red - $e^-$, Blue - $e^+$, Green - $\gamma$ and $E=2E_c$}
\label{e-1MeV-3}
\end{center}
\end{figure}

If one considers an electron from cosmic ray secondaries with energy $\varepsilon<\varepsilon_c$ flying vertically into world volume from the bottom then one can see what any avalanche in the considered volume is not created and as expected the electron perform several hits with an air atoms but rapidly loses his energy down to zero. In the opposite case with energy obeys RREA criteria $\varepsilon = 1 MeV > \varepsilon_c$ which flux is $\sim 10^3 m^{-2} s^{-1}$ one instead obtains an avalanche shown in figure \ref{e-1MeV-3}.

Figure \ref{e-1MeV-3} shown a complicated structure of the cascade of particles and demonstrates general features of RREA: exponential growth of electron number density with characteristic length $l_a$ which is  $\sim 100 $ m near Earth and the role of gamma quanta which are able to produce both an electron-positron pairs and an electrons due to an ionization what is nicely shown on the right top in figure \ref{e-1MeV-3}.

\begin{figure}[h!]
\begin{center}
\includegraphics[width=3.6 in]{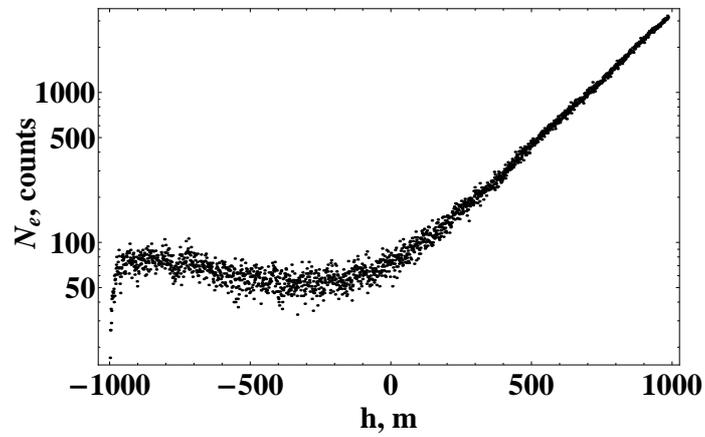}
\caption{The growth of an $e^-$ number ($\varepsilon > 1 MeV$) with height in the RREA. $h=0$ correspond to the center of the world volume}
\label{Ne-h_bb1MeV}
\end{center}
\end{figure}

The corresponding growth of an electrons number with height in RREA
for more mature stage is demonstrated in figure \ref{Ne-h_bb1MeV} where electrons with energy $\varepsilon > 1 MeV$ are considered.

\begin{figure}[h!]
\begin{center}
\includegraphics[width=3.6 in]{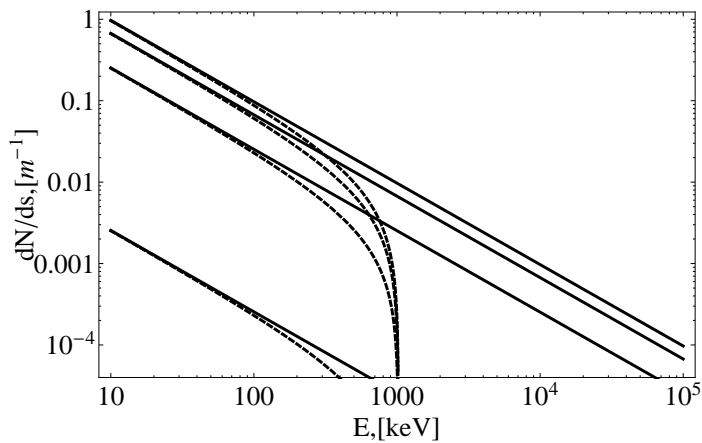}
\caption{Number of ionization $e^-$'s per unit length on different altitudes $h$. Solid line - particle with $\varepsilon >>m_e c^2$, dashed - $\varepsilon\sim m_e c^2$, $h=0,3,10^4, 3\times 10^4$ m from top to bottom respectively}
\label{N_ioniz_electrons_h}
\end{center}
\end{figure}

Remind that this cascade originates from single primary particle, which in turn is able to generate free electrons and the respective rate of particle production is shown in figure
\ref{N_ioniz_electrons_h}.
Obviously, some of the new particles  generated according figure \ref{N_ioniz_electrons_h} can become the seeds for development new RREAs.

\begin{figure}[h!]
\begin{center}
\begin{tabular}{cc}
\includegraphics[width=3.2in]{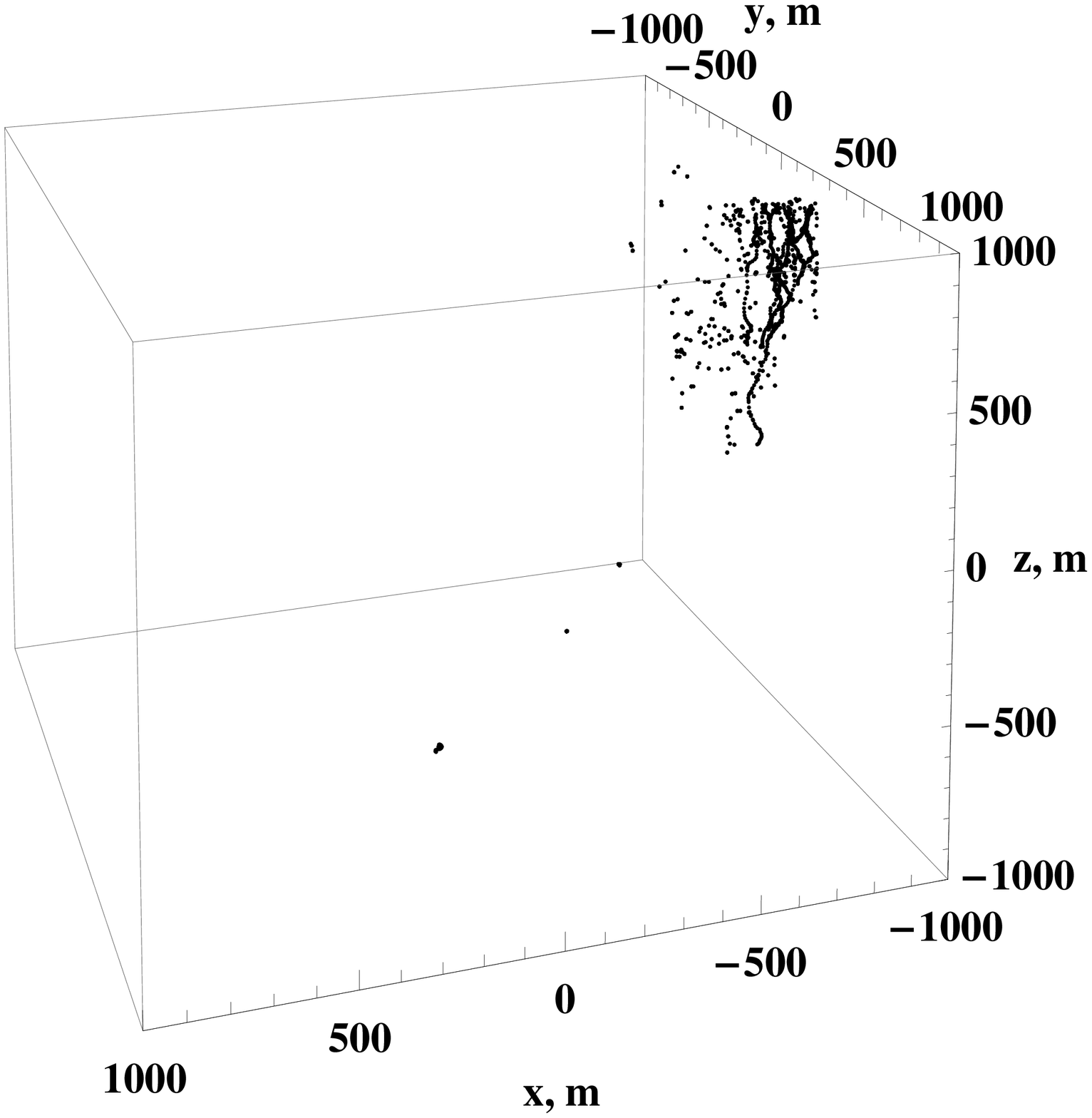} &
\hspace{-1.5cm}\includegraphics[width=3.5in,height=2.8in]
{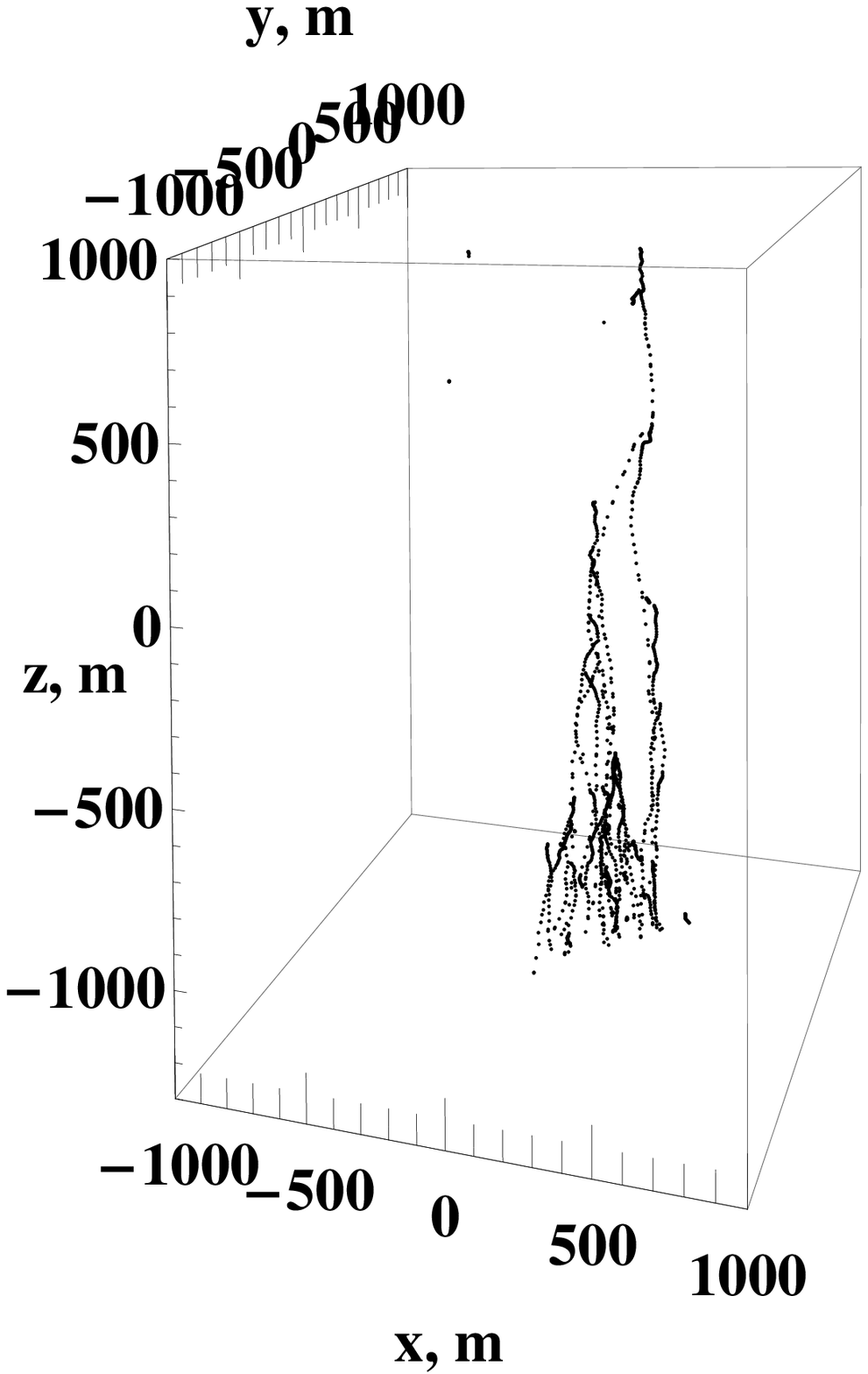}\\
a & b
\end{tabular}
\center\caption{Electrons only distribution. a:  primary particle - $e^+$, $\varepsilon = 30$ MeV, b: primary particle - $e^+$, $\varepsilon = 10$ MeV and with opposite electric field direction -   from the bottom to the top}
\label{TGF3D-pic}
\end{center}
\end{figure}

It is interesting to compare different physical situation at the  initialization of a cascade.
First, one can send a positron in the wold volume with the same direction of the electric field. If the positron has low enough energy  it immediately stops due to both slowdown of the field and ionization loses. If the positron's energy is higher it more effectively produces photons which are able to generate a new RREA. Realization of this situation is shown in figure \ref{TGF3D-pic} a.
The same picture appears if photons generate a RREA cascade like primary particles directly.

Second, one can inverse the field direction and send a positron to be accelerated. Then the positron propagating in the media gains energy enough to more effective photoproduction and ionization,  generating, thus, return avalanche which is shown in figure \ref{TGF3D-pic} b.

\begin{figure}[h!]
\begin{center}
\includegraphics[width=3.2 in,height=3.3in]{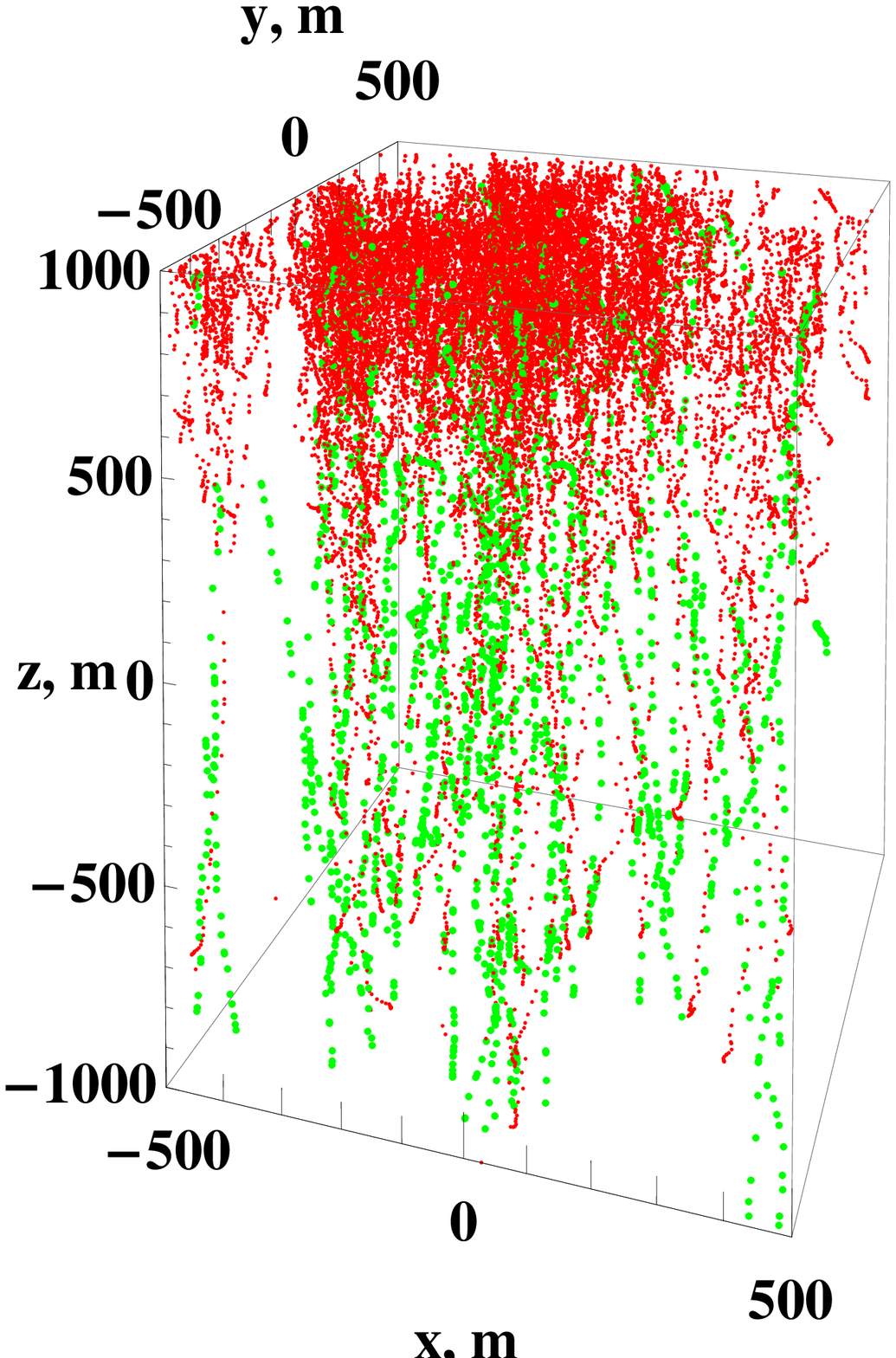}
\caption{\mbox{Particles cascade on more mature stage. Green - $e^+$, Red - $e^-$}}
\label{TGF_tracksEl+Poz}
\end{center}
\end{figure}

A combination of both effects of upward electron avalanches and downward positron stream together with respective photoproduction  due to a bremsstrahlung
establish a feedback chain because of a positron
in turn is able to produce new photons and to initiate thus new avalanche on the lower height what correspond to the inverted figure  \ref{TGF3D-pic} b\footnote{\mbox{The importance of this feedback between different particles in a cascade  was mentioned recently in \cite{2012feedback}}.}.

As result one obtains a series of cascades which increase energetic particles number density significantly as it is shown in figure \ref{TGF_tracksEl+Poz}.

\begin{figure}[h!]
\begin{center}
\includegraphics[width=3.5 in]{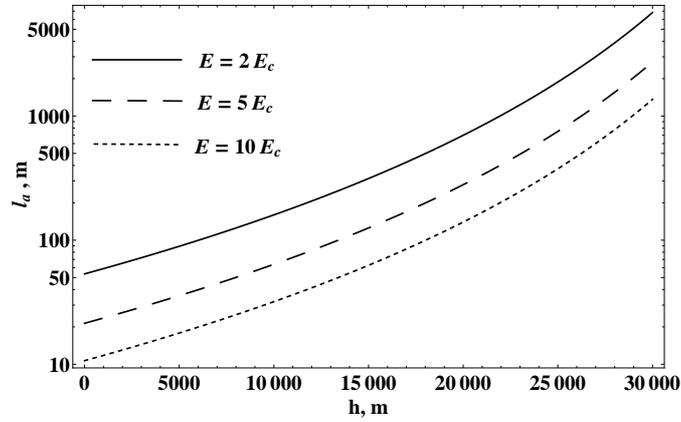}
\caption{Length scale for generation air runaway electrons avalanche}
\label{la}
\end{center}
\end{figure}
Remind, what we traced here mostly formation and initial stage of RREA corresponding to $\Delta h=2000 \,m$.
The typical length scale of a RREA breakdown generated in upper atmosphere extends up to ionospheric height thus $\Delta h\simeq 50\, km$ and for RREA breakdown starting above thunderclouds at $h\approx 20 \, km$ one gets $\Delta h/l_a\geq 20\div40$ where the dependence $l_a(h)$ shown in figure \ref{la} is included. The quasistatic electric field appeared immediately after +CG discharge has
relaxation time $\sim few $ seconds.
This indicates about the huge number of produced particles $N \sim \exp(\Delta h/l_a)\sim 10^{20}$ \cite{gurevich-ufn}.

\begin{figure}[h!]
\begin{center}
\begin{tabular}{cc}
\hspace{-.5cm}\includegraphics[width=3.25in]{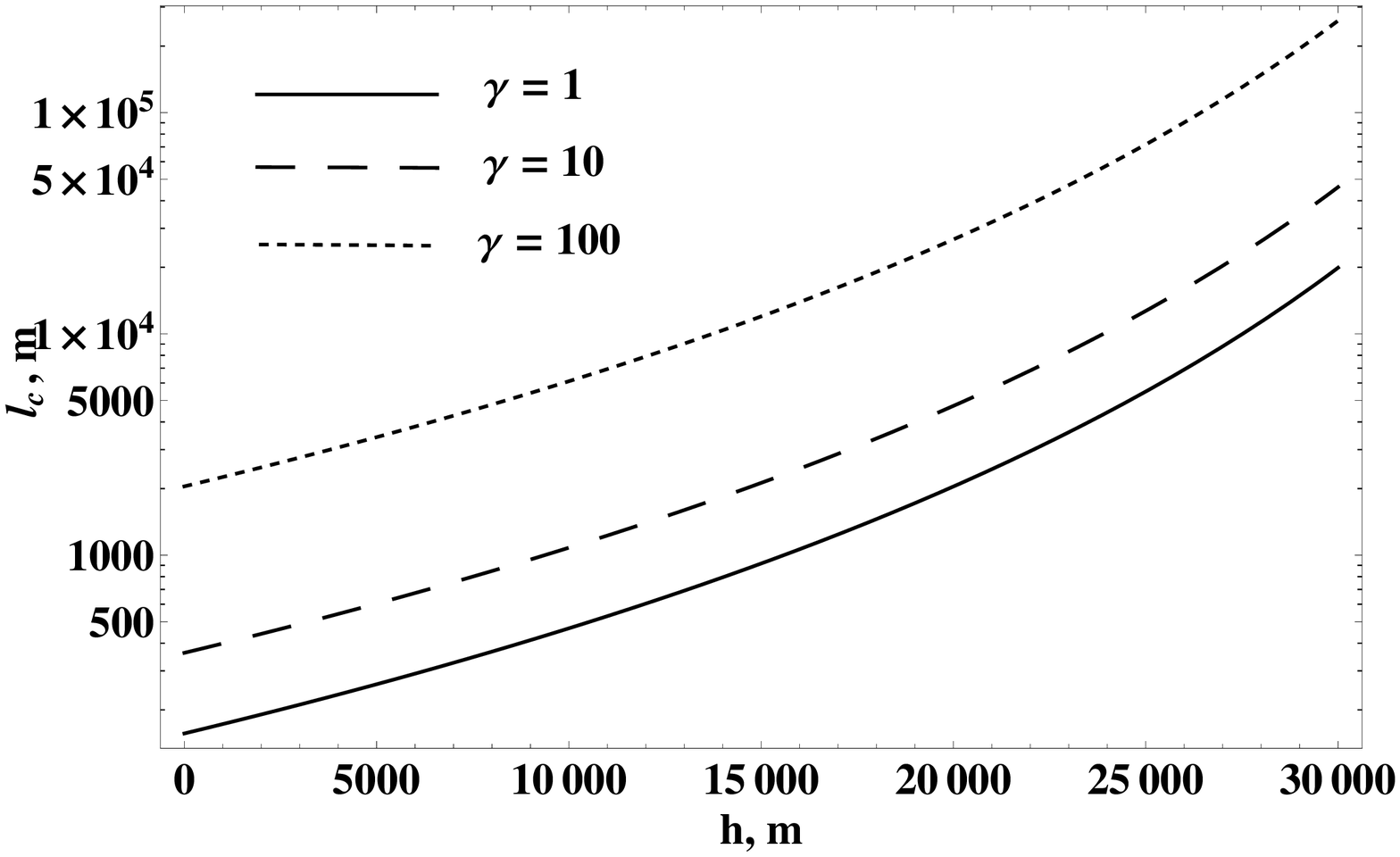} &
\hspace{0.5cm}\includegraphics[width=3.25in]{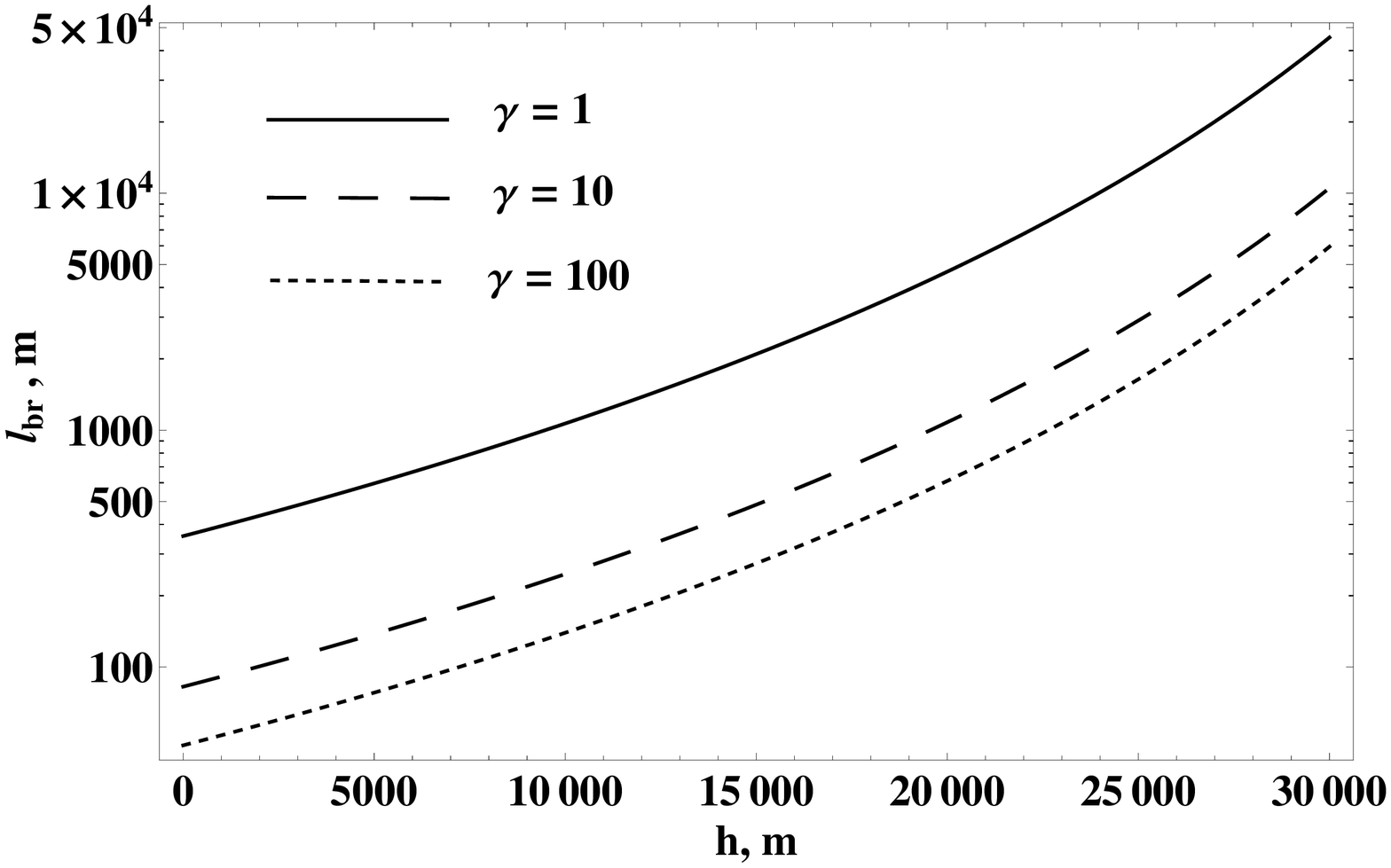}\\
a & b
\end{tabular}
\center\caption{Main free path of photons due to Compton scattering (a), where $\gamma=\varepsilon/m_e c^2$ and electrons (b) due to bremsstrahlung with respect to height in air}
\label{lc-lbr}
\end{center}
\end{figure}

It is interesting to compare $l_a$ with main free path of particles because of their interactions what is shown in figures \ref{lc-lbr}. 

\vspace{-1.0 cm}

One can see what Compton scattering of photons at considering height $h\simeq 5 \, km$ is comparable with a characteristic length scale of RREA generation $l_a$ for small energies $\gamma \sim 1$ and small field $E\sim 2 E_c$. More energetic photons are less liable for interactions and for instance if $\gamma \sim 100$ they escape world volume without interactions. A bremsstrahlung increases with energy instead and leads to more effective photoproduction comparable with $l_a$ at $h\simeq 5 \, km$ at high energies. Obviously both effects recede at high altitudes and particles became free after generation at lower atmosphere. One can estimate what at $h\simeq 30 \, km$ electrons generated with no so high energies $\gamma \lesssim 10$ are free particles with respect to bremsstrahlung and photons propagate without collisions at $\gamma \gtrsim 10$.

\section{Hard spectrum of TGF}
\label{hard-sp-toc}

\begin{figure}[h!]
\begin{center}
\includegraphics[width=4 in]{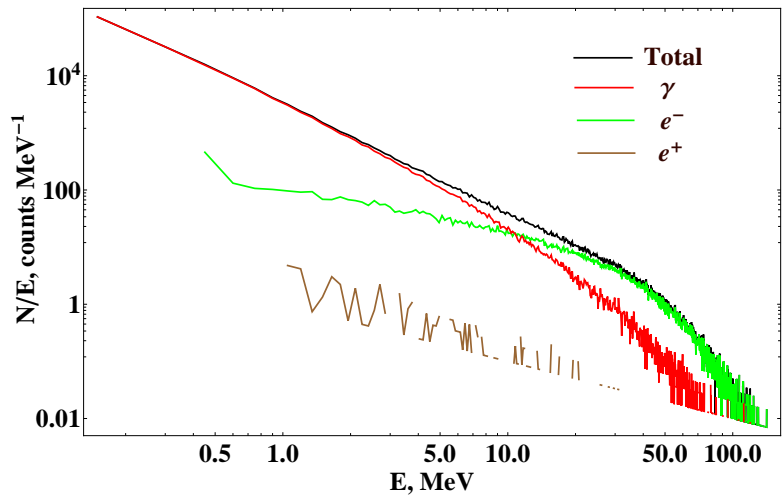}
\caption{Energetic spectrum of typical runaway cascade}
\label{Spectra_ALL}
\end{center}
\end{figure}

Performing simulation with sufficient statistic allow to obtain a spectra of different components of a RREA which is shown in figure
\ref{Spectra_ALL}. One can see what total energety spectrum forms mostly by photons at low energy $\varepsilon \lesssim 10\, MeV$ and by relativistic electrons in the opposite case. Positron flux is sufficiently suppressed at the top of modeling volume which is obvious in the framework of our simulation set up where positrons always directed downward.
Thus from inner parts of atmosphere
upward onto lower density slices a bremsstrahlung photons and a runaway electrons are escaped. According to the simulation the positrons moving along the electric field have a little chance to be distributed in the higher atmosphere. Thus experimental evidence of positrons space detection  \cite{el-pos} implies that their formation is located on the regions without electric field and what they are produced by a huge number TGF cascade originated at the lower altitude where electric field amplitude is sufficient to accelerate an electron beam to produce this TGF as it is clear from interactions shown in figure \ref{lc-lbr}. Thus clear what TGF initiates in the lower atmosphere but where specifically is the matter of discussion
\cite{xu2012source}. Difficulties come from peculiarities of electric field generation, its dynamics and influence on the progress of RREA.
Practically all the  TGF generation models exploit  the idea of RREA, but apply to different field configuration. For instance alternatively to the previous mechanism, TGF can be generated by a  strong electromagnetic pulse produced by a lightning discharge \cite{inan2005production}. Differences in these models results in  varying TGF origination height of about several kilometers.

Our modeling ascertains general picture of TGF and its relation to the RREA. Now we are not addressing to the problem of TGF origination  but look for features of the hard gamma spectrum production. Namely, the question is about an existence of a high-energy spectral component in addition to the well-known power low component extending up to $\sim$ 10 MeV. The flux dependence $F(\varepsilon)\sim \varepsilon^{-2.7\pm0.1}$ for the hard component were obtained together with the registration of photons with energy up to 100  MeV \cite{PRL_AGILE_11}. The problem is that there is no reasonable explanation of such hard spectrum component because of the prediction of usual TGF models according to which  a "power-law spectrum with
an exponential cutoff near 10  MeV is expected with characteristics that are quite independent of the conditions" \cite{PRL_AGILE_11}.
\begin{figure}[h!]
\begin{center}
\includegraphics[width=4 in]{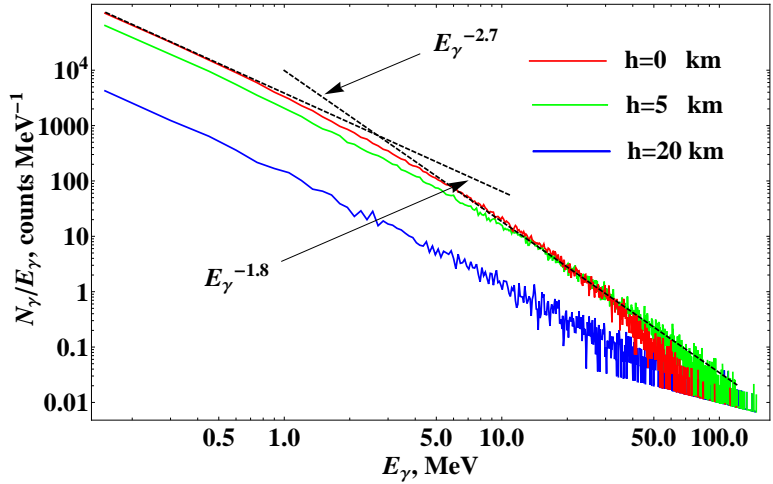}
\caption{Spectrum of TGF produced on different heights and its low and hard asymptotics. h correspond to the position of world volume center}
\label{Sp_gamma_2}
\end{center}
\end{figure}

Our modeling shows some remarkable features of gamma-ray production in RREA. First, it demonstrates an absence of such exponential cutoff in the photon spectra as it is according figure
\ref{Spectra_ALL} at list for our simple set up. Second, where are hard photons at energies about 100 MeV. Third, one can see an electron and a photon dominance changes in the spectrum approximately near 10 MeV which suggests that the hard electrons with $\varepsilon > 10 $ MeV may be responsible for the new additional hard gamma spectral component. Moreover, one can check how photon spectra forms at different altitudes and conclude what according to figure \ref{Sp_gamma_2} low energy branch has approximately the same inclination for all the considering heights. But high energy branch change its inclination with altitudes and near $h \gtrsim 5$ is close to the observed value $-2.7$. Generally speaking the spectral index $s$ of a hard gamma spectrum $F_\gamma\sim \varepsilon_\gamma^{-s}$ decrease with height trying to flatten the entire spectrum.
Low energy limit $s\thickapprox 1.8$ differs from $s\thickapprox 0.5$ observed in \cite{PRL_AGILE_11} and indicates what low energy photons are more intense absorbed and scattered (see figure \ref{lc-lbr}), thus to improve results one needs to regard scale as long as the entire atmosphere.

Taking into consideration huge number of produced particles in RREA breakdown, let us check a possibility that source of generation high energy photons is Comptonization on relativistic electrons of RREA.
Namely, one can regard a problem about an atmospheric tube filled with energetic electrons with increasing number density. We have a lot of radiation in the volume and interesting about the transformation of the radiation passing through this volume due to scattering by the high electron gas inside the tube.

Strictly speaking one needs to consider a full kinetic problem in the manner of Kompaneets equation for energetic photons interacting with ultrarelativistic  electrons and is the task for additional investigation.

Now we restrict ourselves by a simple situation. Namely, the simulation presented here allows immediately to check the possibility of formation hard spectrum of TGF at the limits of single inverse Compton scattering on ulrtarelativistic electrons. It is known what for electron flux $F(\varepsilon_e) \,= \,C \varepsilon_e^{-p}$
where $C=const$, accounting the only one  Compton scattering results in transformation of initially monochromatic photon spectra to the form looks like synchrotron emission: $F_\gamma\sim \varepsilon_\gamma^{-s}$, where $s=(p-1)/2$ \cite{rybicki2008radiative}.
Thus if hard  photons with $s=2.7$ produced due to this mechanism then spectrum of an electrons will be with $p=2s+1$ and we obtain something like $p\simeq 6.4$ what obviously disagree with electrons RREA spectra which according to figure \ref{Spectra_ALL} at high energy goes approximately parallel to the photons one and thus should be $p\simeq s = 2.7$.
One can state what high energy photons appeared due to Comptonization process may play a negligible role at least in simplest case and what a bremsstrahlung is a basic mechanism for the hard photons production.

\section{Conclusions}

We perform numerical simulation of runaway electron breakdown in the upper atmosphere using GEANT4 simulation toolkit.
General features of RREA are reconstructed and
the feedback which arises in connections between different cascades originated from different particles is emphasized.

Characteristics of radiations accompanying breakdown are obtained.
We can see what at least in the simple case of a $\Delta h \simeq 2 \, km$ and $R=1\, km$ cylinder filled with uniform air atmosphere
and with a thunderstorm's electric field applied
there is no strongly marked exponential cutoff in the photon spectra. Next, at the energy near 10 MeV of experimentally discovered breaking of the photon spectra electrons dominates in the total particle spectrum of relativistic cascade but hard photons at energies about 100 MeV and above are definitely generated.
We demonstrated dependance of the high energy branch of photon spectra with respect to an altitude and find what at the reasonable parameters hard photons have spectral index close to the observed value. This opens up a possibility to establish a new method to  estimate a TGF generation height by future more accurate TFG spectrum identification.

For a detailed investigation of a TGF production and propagation one needs to take into account more realistic properties of the model like dependence of air density on altitude, nonstationarity of an electric field and increase a scale of modeling up to dozens of kilometers when particle interactions became negligible. This conditions can sufficiently increase calculation time and require respective hardware platform.
On the other hand, for comprehensive theoretical analysis of particles interactions and to establish a nature of new hard gamma component in relativistic cascade \cite{PRL_AGILE_11} one needs to
analyze  kinetic equations with collisional integral responsible for all relevant interactions in the manner of
\cite{1998_Gurevich_kin_eq}. The simplest estimation based on spectral shapes  can strongly suppress contribution to the hard proton production at least in the simplest case of a single inverse Compton scattering on relativistic electrons, but detailed analysis is a matter of further investigations.

Finally we can add what confirmation of generation hard photon component in TGF substantiates a reconsideration of possible radiation doses which are able to affect on manned aircrafts
\cite{TGF_hazard}.

\bibliographystyle{unsrt}
\bibliography{bib_TGF}

\end{document}